\documentclass[journal=jacsat,manuscript=article]{achemso}
\usepackage[table]{xcolor}
\usepackage{colortbl}
\usepackage{booktabs}
\definecolor{softblue}{RGB}{220,230,245}
\definecolor{softpink}{RGB}{245,220,230}

\usepackage[version=3]{mhchem}
\usepackage{amssymb}
\usepackage{graphicx}
\usepackage[colorlinks=true,breaklinks=true,linkcolor=black,citecolor=green,urlcolor=magenta]{hyperref}
\usepackage{physics}

\usepackage[table]{xcolor}

\definecolor{PastelBlue}{rgb}{0.68,0.85,0.90}
\definecolor{PastelPink}{rgb}{1.0,0.91,0.95}

\title{A Quantum Multi-Programming Framework to Maximize Quantum Resources for the LUCJ Ansatz}

\author{Milana Bazayeva}
\affiliation{Center for Computational Life Sciences, Lerner Research Institute, The Cleveland Clinic, Cleveland, Ohio 44106, United States}

\author{Abigail McClain Gomez}
\affiliation{IBM Quantum, IBM Research, Cambridge, MA 02142, USA}

\author{Kenneth M. Merz Jr.}
\email{kmerz1@gmail.com}
\affiliation{Center for Computational Life Sciences, Lerner Research Institute, The Cleveland Clinic, Cleveland, Ohio 44106, United States}
\alsoaffiliation{Department of Chemistry, Michigan State University, East Lansing, Michigan 48824, United States}

\begin{document}
\maketitle

\section{Abstract}
In the context of quantum computing, efficient resource management is crucial for optimizing throughput on cloud-based platforms and maximizing hardware utilization. In the present work, we propose an approach to tackle quantum chemistry problems via quantum multi-programming of the Local Unitary Cluster Jastrow (LUCJ) ansätze. The ground-state energy of the molecular system is obtained via Sample-based quantum diagonalization (SQD), further refined by its extended version (ext-SQD). Building upon the Qiskit Experiments package, which already supports parallel execution functionality for general tasks, we developed a novel parallel experiment class tailored for quantum chemistry problems. 
Cross-talk is a known issue in the multi-programming frameworks and can corrupt the ground-energy estimation of the simulated systems. To assess its impact within our approach, we simulated two conformations of the ethanol molecule: one at the equilibrium state (EtOH$_{Eq}$), and one with the O-H bond stretched to 1.2 ${Å}$ (EtOH$_{1.2}$). We defined three different layouts that we executed in a randomized fashion, alternating serial and parallel execution within 10 independent replicates. The single modality of each circuit was kept as a baseline to evaluate the effect of cross-talk induced by quantum multi-programming. The energies obtained at the first-, last- and ext-SQD iteration were compared to the classical Heat-bath Configuration Interaction (HCI) reference. Our findings highlight the viability of a quantum multi-programming workflow for quantum chemistry as the robust post-processing protocol effectively mitigates possible cross-talk induced noise. At the final step of the configuration recovery process, the energy difference relative to the HCI reference is negligible, within 0.001 kcal/mol.

\section{Introduction}

Current pre-fault-tolerant quantum devices marked a milestone in the transition of quantum computing from theory to practice. However, these quantum processing units (QPUs) come with constraints, including limited qubit counts, noisy operational gates, and short decoherence times.\cite{Preskill2018quantumNISQ,Bethel2026,LeCompte,QuantumMapping} Even state-of-the-art architectures, such as the Heron R3 chip with 156 qubits, available through IBM's quantum computing environment, require careful resource management for reliable simulations. As a result, quantum circuit execution is limited by size and depth, making accurate simulations feasible mainly for small- and medium-scale systems. \cite{Bharti,niu2023multiprogramming} Cloud-based quantum computing platforms have lowered the barrier to access quantum hardware. Services like IBM Quantum let users run experiments on QPUs without owning physical devices, and their free plan was recently upgraded to enable 20 minutes per month of device access. \cite{IBMQuantumPlans2026} 
However, even this expanded access comes with practical limitations: the total time required to execute a circuit is the sum of the queueing and execution times; the free monthly quota still restricts non-premium users. These factors highlight the need for more efficient workload execution strategies for pre-fault-tolerant hardware. In this scenario, quantum multi-programming has emerged as a promising approach in which multiple quantum circuits are executed in parallel on the same QPU. \cite{DasMultProg,LiuMultProg,MezzaMottaVQEParallel} By leveraging the QPU's underutilized spatial dimension, multi-programming increases hardware throughput and reduces queue times, at the expense of fewer qubits available for each individual circuit execution. \cite{niu2023multiprogramming} During parallel circuit execution, the proximity of active qubits can potentially raise the likelihood of hardware-level interference, known as cross-talk. \cite{Ohkura2021CrossTalk,murali2020crosstalk}
This phenomenon can be defined as the effect arising from undesired interactions among hardware components, such as electronic controls, cooling systems, and qubit coupling. This violates locality and the independence of gate operations. \cite{SakiMultiProg,Sarovar2020detectingcrosstalk,smultitenant} 

The cross-talk effect can be either operational with interferences occurring during simultaneous operations \cite{PRXQuantum,murali2020crosstalk,Exp-oper,smultitenant}, or idling with changes in the state induced by the neighboring qubits. \cite{idling,smultitenant} Previous studies have focused on the analysis of the effects induced by ZZ interactions \cite{HighCintrastZZ,SuppCross,niu2023multiprogramming} and CNOT gates. To investigate the effect of CNOT-induced cross-talk, Simultaneous Randomized Benchmarking (SRB) \cite{SRB} was developed within the multi-programming paradigm. It was demonstrated that on IBM devices, the cross-talk effect is relevant at a distance of one hop between CNOT pairs \cite{murali2020crosstalk} and weakens with increasing separation. The occurrence of cross-talk can increase gate error rates by an order of magnitude \cite{murali2020crosstalk,niu2023multiprogramming}, which is detrimental to high-precision tasks like chemical ground-state estimation. Previous studies on multi-programming have demonstrated the trade-off between higher throughput and potential fidelity degradation \cite{Davenport2023}, further motivating the present work. While previous studies (such as the Quantum Multi-programming Compiler (QuMC) framework) have proposed cross-talk-aware qubit partitioning and parallelism management, their evaluations focused on generic Variational Quantum Eigensolver (VQE) circuits.\cite{niu2023multiprogramming,MezzaMottaVQEParallel} In contrast, our study targets quantum chemistry workloads, evaluating whether such a paradigm can reliably produce ground-state energies. 

Concurrent execution of multiple circuits is a significant hardware challenge. Limited connectivity and non-uniform qubit quality make it difficult to partition qubits and avoid possible cross-talk. To assess the feasibility of the proposed parallel LUCJ scheme, we perform both parallel and single-circuit simulations and we systematically compare the resulting energies to classical Heat-Bath Configuration Interaction (HCI) \cite{holmes2016heat} reference values. This allows us to evaluate the trade-off between improved execution efficiency and possible accuracy loss from concurrent execution, providing a practical evaluation of parallel workloads for quantum chemistry on current quantum hardware.


\section{Methods and Computational Details}

In the present work, we integrated the execution of the LUCJ anstaz within the native parallel execution infrastructure provided by the Qiskit Experiments package (version 0.13.0) \cite{QiskitExperiments}. To adapt this framework to our chemistry-related problems, we introduced two new experiment classes as described below.

\subsection{The \texttt{LUCJ\_SubExperiment} Class}

The \verb|LUCJ_SubExperiment| class (Figure \ref{fig:Figure1}) is responsible for the generation of parametrized circuits by integrating \textit{ab initio} electronic structure data with the LUCJ ansatz. The user has to provide the FCIDUMP file, which is central in the circuit generation as it contains all the details of the chemical system. 
The one- and two-electron integrals of the orbitals within the active space are extracted and used to establish a Restricted Hartree-Fock (RHF) reference. This information is then passed to coupled cluster singles and doubles (CCSD) calculations\cite{fajen2025accelerating} to obtain the single $t_1$ and double $t_2$ excitation amplitudes. Specifically, the $t_2$ amplitudes, used to construct the UCJ operator within the \verb|UCJOpSpinBalanced.from_t_amplitudes| method via \verb|ffsim| \cite{ffsim2024}, are enhanced by double factorization. The circuit is optimized using the L-BFGS-B \cite{t2opt,OptLUCJ} algorithm in SciPy \cite{SciPy} with a maximum of 1000 iterations. This optimization process minimizes the difference between the original CCSD amplitudes and the compressed representation used in the circuit, focusing on the specific interaction pairs defined for the experiment. The circuit is then constructed in two main steps using the Jordan-Wigner (JW) \cite{JordanWigner1928} mapping. First, the \verb|PrepareHartreeFockJW| function initializes the qubits into the HF reference state. Next, the \verb|UCJOpSpinBalancedJW| function applies the optimized LUCJ operator to the qubits. This process translates the fermionic wavefunction into a sequence of quantum gates, forming a single-repetition circuit (\verb|nreps=1|). 

For the hardware execution, the \verb|LUCJ_SubExperiment| class requires a list of physical qubits for each sub-experiment. These layouts must allow the circuit mapping onto disjoint partitions of the QPU to avoid cross-talk effects by ensuring at least one layer of idle qubits. The ancillas are defined as an external parameter and can be customized for each sub-experiment. To ensure a seamless identification and retrieval of the counts dictionaries, the user must provide a unique name for each sub-experiment that is stored in the metadata. In the present work, we manually partitioned the qubits for each sub-experiment, but the proposed framework can support automatic qubit selection strategies.
By bundling these hardware and software parameters together for each sub-experiment, we ensure that even complex, multi-molecule experiments remain tractable and reproducible.

\subsection{The \texttt{LUCJ\_ParallelExperiment} Class}

\begin{figure}
    \centering
    \includegraphics[width=1\linewidth]{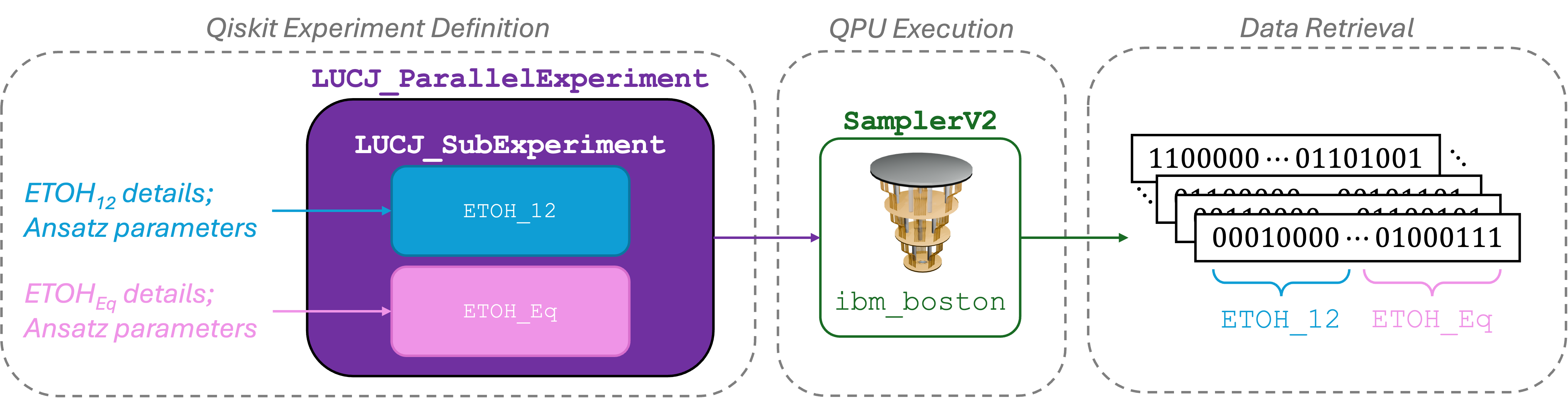}
    \caption{Schematic representation of the quantum multi-programming workflow for the LUCJ ansatz. First, the experiment is constructed from individual sub-experiments. The final \texttt{LUCJ\_ParallelExperiment} is executed on the QPU (\texttt{ibm\_boston} for all results presented here). Finally, the data is retrieved and split according to the underlying sub-experiments.}
    \label{fig:Figure1}
\end{figure}

The \verb|LUCJ_ParallelExperiment| class acts as an orchestration layer (Figure \ref{fig:Figure1}), managing multiple \verb|LUCJ_SubExperiment| instances and merging them into a single execution job. Each sub-experiment is transpiled independently using an optimized \verb|PassManager| generated through \verb|generate_preset_pass_manager| at optimization level 0. This choice ensures that the circuit topology and qubit mapping remain unchanged after transpilation, preventing the compiler from using qubits outside of the initial selection. The process uses a specific approach from the \verb|ffsim| package. The \verb|PRE_INIT| passes from the \verb|ffsim| package are used to decompose orbital rotations into hardware-compatible gates. To minimize circuit depth and eliminate redundant operations, we introduced a post-initialization stage using the \verb|RemoveIdentityEquivalent| pass, which deletes the near-identity rotations. Additionally, we utilized the \verb|FoldRzzAngle| pass to further condense the interaction gate sequences, optimizing the final execution on the QPU. Each resulting transpiled circuit is passed to the \verb|compose| method, which generates a single final circuit. Simultaneously, the \verb|ClassicalRegister| of each sub-experiment is extracted and inserted into the global circuit. This ensures that each molecule maintains its own dedicated measurement container, preserving the independence of the data upon parallel execution. In addition, the \verb|measure| method is applied to couple the specific \verb|physical_qubits| of each sub-experiment directly onto its corresponding \verb|ClassicalRegister| object. By doing so, the bitstrings generated by the hardware are stored in the correct register, which is subsequently identified in the post-processing phase using the unique labels stored in the metadata. 

\subsection{Sample-based Quantum Diagonalization (SQD) and its extended form (ext-SQD)}

The application of quantum computing to solve chemistry-related problems requires a tailored encoding approach. The LUCJ ansatz was proposed by Motta  et al. \cite{Motta2023LUCJ}, and has the following truncated form:
\begin{align}
|\Phi_{\mathrm{qc}}\rangle =  e^{\hat{K}_1} e^{i\hat{J}_1} e^{-\hat{K}_1} |x_{\mathrm{RHF}}\rangle
\end{align}
where the density-density operator is represented by $ \hat{J}_1 $, whereas $ \hat{K}_1 $ and $ \hat{K}_2$ correspond to the one-body operator. Finally, the last term of the equation represents the RHF bitstring where \( \mathbf{x} \in \{0,1\}^{2M} \). The transition from $M$ spatial to $2M$ spin orbitals is performed by the JW transform, allowing the encoding and mapping of the initial wavefunction onto physical qubits. This scheme imposes locality on the $ \hat{J}_1 $ operator, and mediates the density-density interactions across opposite spin states through ancillary qubits, resulting in the zig-zag layout (a standard layout depiction is given in Figure S1). This localized circuit architecture bypasses the all-to-all connectivity which is hard to achieve on heavy-hex devices and requires limited SWAP gates. However, because of the intrinsic flexibility of the underlying ansatz, we explored slightly different layouts to accommodate multiple circuits on a QPU with a limited qubit count. 

From a practical point of view, the amplitudes used to parametrize the LUCJ circuit are extracted from gas-phase restricted closed-shell CCSD calculations within the selected active space, as done in previous quantum-centric studies and better described in the \verb|LUCJ_SubExperiment| section. The computation of the electronic ground states using pre-fault-tolerant devices is challenged by decoherence and gate errors, which often lead to sampled configurations that violate physical symmetries. To mitigate these effects, we used the SQD algorithm \cite{robledo2024chemistry,kaliakin2024accurate}; a self-iterative procedure that extracts only the configurations contributing the most to the ground-state energy.
The circuit execution and the repeated measurements (quantum shots) of the state $\ket{\Psi}$ generates a set of bitstrings $\tilde{\chi} = \{ \mathbf{x} \mid \mathbf{x} \sim \tilde{P}_\Psi(\mathbf{x}) \}$, where \( \mathbf{x} \in \{0,1\}^M \). 

Each bitstring represents an electronic configuration in the computational basis, the Slater determinant. The recovery process starts by scanning $\widetilde{\chi}$ for entries with unphysical states. In the context of a second-quantization mapping like JW, bitstrings corresponding to unphysical states will violate particle conservation by having an incorrect Hamming weight. The recovery process applies a probabilistic bit-flip based on the discrepancy between the current bits and the current estimate of the average spin-orbital occupancies $p\sigma$. For the initial iteration, the occupancies are directly estimated from the raw samples that correspond to physical states. This step generates a set of recovered configurations $X_R$ that is partitioned into $K$ batches $S(1), \ldots, S(K)$. Each batch represents a subspace ($S^{(b)}$) of the total Hilbert space over which the Hamiltonian $\hat{H}$ is projected by the projector:
\begin{equation}
\hat{P}_{S^{(b)}} = \sum_{x \in S^{(b)}} |x\rangle \langle x|
\end{equation}
The projected Hamiltonian form is:
\begin{equation}
\hat{H}_{S^{(b)}} = \hat{P}_{S^{(b)}} \hat{H} \hat{P}_{S^{(b)}} 
\end{equation}

By diagonalizing these projected operators, we obtain a collection of estimates of the local ground state $|\psi^{(b)}\rangle$ and its energy $E^{(b)}$. The estimate of lowest energy is used to update the orbital occupancies via:
\begin{equation}
n_{p\sigma} = \frac{1}{K} \sum_{b=1}^{K} \langle \psi^{(b)} | \hat{n}_{p\sigma} | \psi^{(b)} \rangle
\end{equation}
The updated $n_{p\sigma}$ serves as an input to the subsequent iteration of the recovery cycle. The process is repeated until convergence is reached.
To further refine the calculation, the lowest energy batch $S^{(b)} \subset \chi_R$ extracted from the last-SQD iteration is optimized through a single ext-SQD step \cite{ext-sqd,barroca2025surface}.
A single-excitation operator $\hat{E}_I$ (via PyCI package \cite{richer2024pyci}) is applied to the $S^{(b)}$ leading to an expanded set of configurations in the form:
\begin{equation}
S_E^{(b)} = \{\, |x\rangle,\; \hat{E}_I |x\rangle \,\}.
\end{equation}
The projector on the augmented space is defined as:
\begin{equation}\hat{P}_{S_E^{(b)}} 
= 
\sum_{z \in S_E^{(b)}} |z\rangle \langle z|,
\end{equation}
and the resulting Hamiltonian is defined as:
\begin{equation}\hat{H}_{S_E^{(b)}} = \hat{P}_{S_E^{(b)}} \hat{H} \hat{P}_{S_E^{(b)}}.\end{equation}
The final energy is obtained by diagonalizing $\hat{H}_{S_E^{(b)}}$. This approach explores additional configurations, improving the accuracy without introducing significant computational overhead \cite{ext-sqd}.
In our workflow, we employed the Qiskit Addon: SQD (version 0.12.0) \cite{sqdaddon} to perform the configuration recovery. We configured the SQD procedure with 10 batches of 3000 samples each and executed it for 5 iterations. At each step, the relevant configurations (carryover threshold of $10^{-4}$) were included in the generated batches. A  threshold of  $10^{-8}$ and $10^{-5}$ was used for the convergence of the total energy and the average orbital occupancy, respectively. The selected basis diagonalization (SBD) solver enabled parallel diagonalization of the subspaces on 48 CPUs. For the final ext-SQD step, we identified the dominant configurations from the lowest energy batch using a more stringent configuration interaction (CI) threshold of $10^{-5}$.  This set of configurations served as the basis for subsequent subspace expansion. 


\subsection{Experiment Execution}

\begin{figure}
    \centering
    \includegraphics[width=1.0\linewidth]{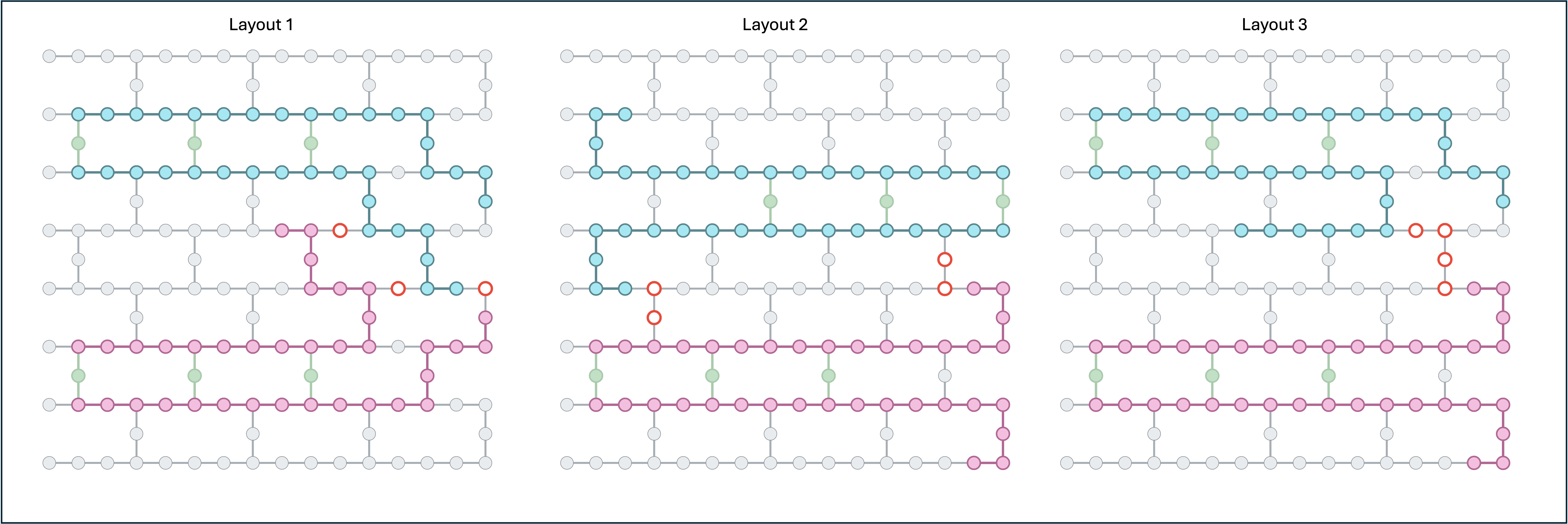}
    \caption{Representation of the three layouts used for the LUCJ-multi-programming runs. The same layouts were also used to run the single experiment as a baseline for the parallel execution. The minimum number of buffer qubits between the two circuits is circled in red. The circuit in blue correspons to EtOH$_{1.2}$, while EtOH$_{Eq}$ is shown in pink; ancilla qubits are reported in green. Within each circuit, blue and pink is used for both spin orbitals. The different spin sectors are not distinguished in the figure, as the colors only differentiate the two circuits.}
    \label{fig:layouts}
\end{figure}

To ensure statistical robustness and to account for hardware noise variability during the experimental execution, we generated three different multi-programming layouts (Figure \ref{fig:layouts}). In addition, the simulations were performed under the Randomized Block Design (RBD) scheme: an initial randomization was implemented in the execution layout order, followed by the shuffle of the modalities within each layout. These safety measures were taken to minimize the impact of device temporal drifts and biases in the QPU scheduling. This also justifies the choice of running each replicate (a total of 10) in Batch execution mode through the Qiskit Runtime \verb|SamplerV2| primitive, minimizing the circuit execution delay. To assess the extent of cross-talk within the LUCJ-multi-programming framework, we initially extracted the energy estimation from the simultaneous execution and compared it to the respective serial run, that served as a baseline. This comparison determined the applicability of the LUCJ-multi-programming paradigm to solve chemistry-related problems while efficiently leveraging computational resources. In the present study, we used the ethanol molecule in two distinct configurations as a benchmark: the equilibrium geometry, EtOH$_{Eq}$, and a configuration with the stretched O-H bond (1.2 ${Å}$), EtOH$_{1.2}$. Notably, these two states are separated by a relative energy difference of approximately 14 kcal/mol, providing a distinguishable baseline to discriminate between the computed energies. Moreover, their energetic estimation through quantum computing showed an increase in accuracy in contrast to uniform sampling, making it a suitable and non-trivial test case. Each proposed layout (Figure \ref{fig:layouts}) is characterized by a different number of buffer idle qubits between the active sub-experiments. All circuits were simulated using three ancillary qubits, as they provide a good trade-off between energy accuracy and layout flexibility. In multi-programming modality, the qubit count is reduced by the accommodation of multiple layouts on the same QPU, while avoiding faulty qubits and gates. Details on ancillary benchmarking are reported in the the Supplementary Information in Figure S1 and Table S1. All runs were executed using an identical parameter set-up: 200,000 quantum shots with the XY4 Dynamic Decoupling (DD) sequence activated to suppress decoherence on the idle buffer qubits during the parallel runs, and suggested in previous works. 

\section{Results and Discussion}
\begin{figure}[H]
    \centering
    \includegraphics[width=1\linewidth]{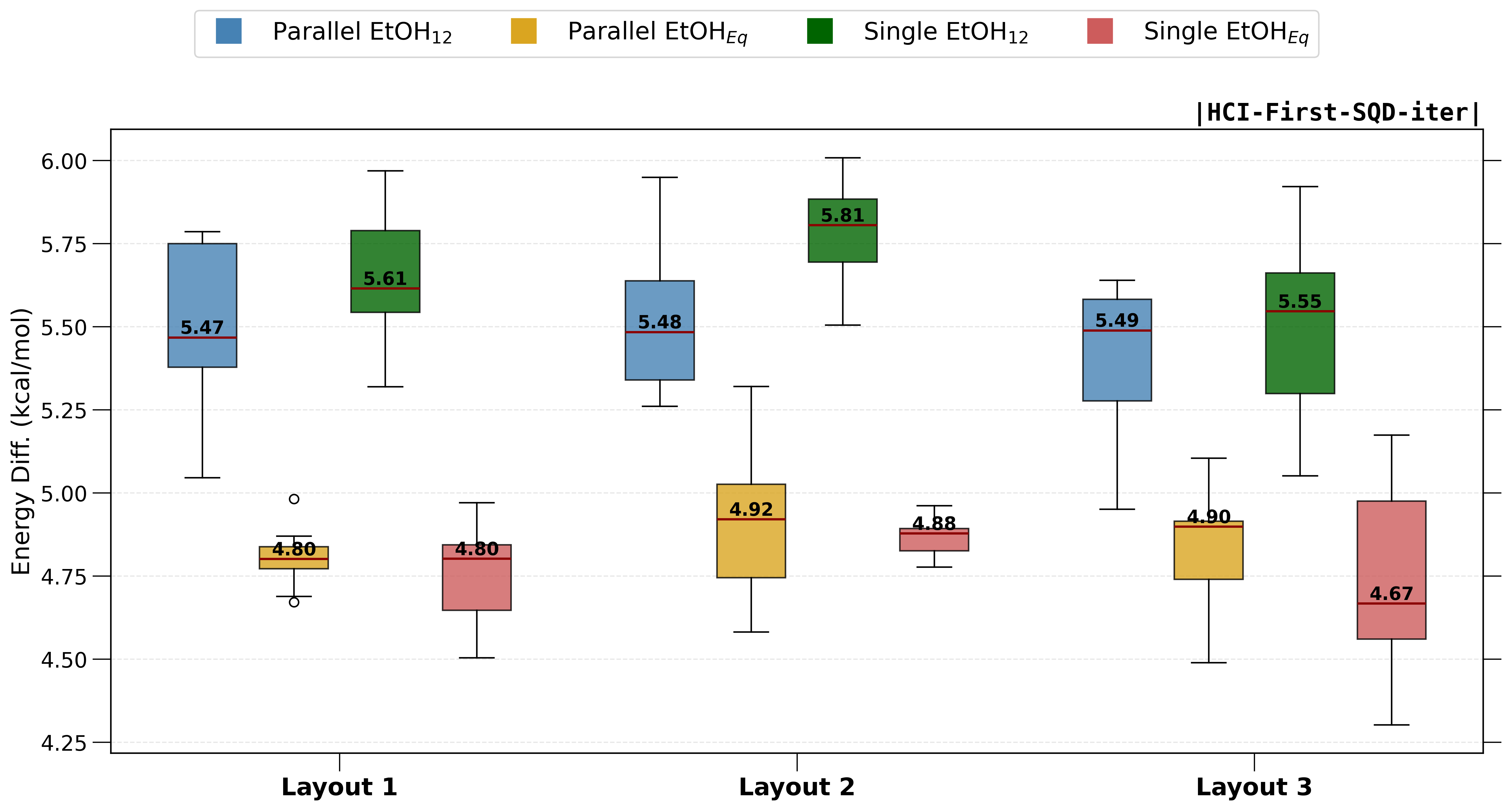}
    \caption{Distributions of the absolute energy deviation from the HCI reference at the First-SQD-iter stage for all layouts and execution modalities, computed across 10 independent replicates. The boxes represent the interquartile range (IQR), with the median shown as a horizontal red line and its value reported above in kcal/mol. The whiskers extend to 1.5×IQR, and the circles denote outliers.}
    \label{fig:Figure3}
\end{figure}

Benchmarking the ground-state energy of a molecular system on quantum hardware poses many challenges and is the result of multiple stochastic contributions: the execution on the QPU itself, followed by the classical post-processing, SQD/ext-SQD. Even for the single circuit and single geometry case, multiple independent runs produce a distribution of energy values rather than a single outcome. In this study, we introduced two additional layers of complexity: first, a challenging system, EtOH$_{1.2}$, whose QPU and SQD/ext-SQD sampling are significantly more demanding than for the equilibrium counterpart; secondly, parallel execution can potentially introduce an additional source of noise through cross-talk. To account for these sources of fluctuation and guarantee statistically reliable conclusions, we collected 10 independent replicates for each experimental configuration. 
The box plot in Figure  \ref{fig:Figure3} is better suited to represent the stochastic nature of the approaches used. It shows the values distribution (reported as  absolute deviation from the HCI reference in the first-SQD-iter stage) across all the replicates and highlights outliers, while also providing the median, which is not affected by extreme values. On the other hand, the mean is an outlier-sensitive metric that can potentially pinpoint atypical trends and can also capture the improvements of the SQD refinement steps.  
Although this initial stage incorporates a preliminary bit-flip correction to restore valid string configurations (as discussed above, in the Methods section), it represents the most minimally processed data available and the best probe for isolating cross-talk-induced energy deviations. 
For EtOH$_{Eq}$, the distributions show comparable medians across all three layouts and execution modes, with no systematic offset that scales with buffer size (Figure \ref{fig:Figure3}). In Layout 1, the parallel and serial execution mode yield essentially identical mean deviations $4.80 \pm 0.09$  and $4.76 \pm 0.14$ kcal/mol, respectively. A similar agreement is observed for Layout 2 ($4.92 \pm 0.23$ vs $4.86 \pm 0.06$ kcal/mol) and Layout 3 ($4.82 \pm 0.20$ vs $4.73 \pm 0.30$ kcal/mol), with all parallel–serial pairs remaining compatible. The change in the distributions from layout to layout does not show a monotonic trend: the first experiment shows outlier values in both energy directions in the parallel mode, Layout 2 has a broader interquartile range for the parallel protocol, and the last configuration displays the opposite pattern, with the distribution obtained from the single execution being wider. Despite the lack of a clear trend across all the layouts in relationship with the different buffer sizes, it does not rule out cross-talk which is arguably present in our samples but overshadowed by the stochastic fluctuations of the employed methods. 
Nonetheless, maintaining a minimal buffer of a single qubit is good practice as suggested by the studies on which our experimental design is based. \cite{SakiMultiProg,Bravyi_2021,murali2020crosstalk}

For EtOH$_{1.2}$, the distributions are markedly broader compared to the equilibrium geometry, consistent with the increased sampling complexity of the stretched O–H geometry at 3K samples per batch. Across all three layouts, the serial execution yields a higher median or comparable values to the parallel mode, counter intuitive to what systematic cross-talk would induce. Layout 1 has a parallel mean deviation of $5.50 \pm 0.25$ kcal/mol, which is lower than the serial counterpart ($5.65 \pm 0.19$ kcal/mol). A similar trend is also observed in Layout 2, with values of $5.53 \pm 0.23$ and $5.79 \pm 0.16$ for parallel and single runs, respectively. The remaining layout shows values that are close, with $5.39 \pm 0.26$ kcal/mol and $5.47 \pm 0.28$ kcal/mol for parallel and single runs. A comparison of the energy distributions as a function of SQD sample size is provided in the Supplementary Information (Figure S2), where batch-size-influenced behavior for this geometry is illustrated at 1K, 3K, and 5K sampling budgets. The 3K sample size was chosen to maintain consistency with the EtOH$_{Eq}$ benchmark and represents a challenging but realistic sampling condition.
\begin{table}[H] 
\centering
\begin{tabular}{lccc} 
\toprule
Layout & $|$HCl – First-SQD$|$ & $|$HCl – Last-SQD$|$ & $|$HCl – ext-SQD$|$ \\
\midrule
\multicolumn{4}{c}{\textbf{Layout 1}} \\
\midrule
Parallel-EtOH$_{1.2}$& $5.50 \pm 0.25$ & $0.56 \pm 0.03$ & $0.152 \pm 0.002$ \\
Parallel-EtOH$_{Eq}$& $4.80 \pm 0.09$ & $0.49 \pm 0.01$ & $0.164 \pm 0.003$ \\
Single-EtOH$_{1.2}$& $5.65 \pm 0.19$ & $0.57 \pm 0.03$ & $0.152 \pm 0.002$ \\
Single-EtOH$_{Eq}$& $4.76 \pm 0.14$ & $0.50 \pm 0.01$ & $0.163 \pm 0.001$ \\
\midrule
\multicolumn{4}{c}{\textbf{Layout 2}} \\
\midrule
Parallel-EtOH$_{1.2}$& $5.53 \pm 0.23$ & $0.47 \pm 0.02$ & $0.148 \pm 0.004$ \\
Parallel-EtOH$_{Eq}$& $4.92 \pm 0.23$ & $0.50 \pm 0.01$ & $0.164 \pm 0.001$ \\
Single-EtOH$_{1.2}$& $5.79 \pm 0.16$ & $0.49 \pm 0.01$ & $0.148 \pm 0.004$ \\
Single-EtOH$_{Eq}$& $4.86 \pm 0.06$ & $0.50 \pm 0.01$ & $0.164 \pm 0.002$ \\
\midrule
\multicolumn{4}{c}{\textbf{Layout 3}} \\
\midrule
Parallel-EtOH$_{1.2}$& $5.39 \pm 0.26$ & $0.58 \pm 0.02$ & $0.151 \pm 0.004$ \\
Parallel-EtOH$_{Eq}$& $4.82 \pm 0.20$ & $0.50 \pm 0.01$ & $0.164 \pm 0.002$ \\
Single-EtOH$_{1.2}$& $5.47 \pm 0.28$ & $0.58 \pm 0.02$ & $0.151 \pm 0.003$ \\
Single-EtOH$_{Eq}$& $4.73 \pm 0.30$ & $0.50 \pm 0.02$ & $0.163 \pm 0.004$ \\
\bottomrule
\end{tabular}
\caption{Mean absolute energy deviation (kcal/mol) from the HCI reference ($\pm$ standard deviation) computed across 10 independent replicates for each layout and execution modality. The energies were extracted at three checkpoints: the first-, last-, and ext-SQD step.} 
\label{tab:hcl_sqd_clean}
\end{table}

The presented results (Table \ref{tab:hcl_sqd_clean}) show that the first-SQD iteration captures both the impact of quantum hardware sampling and the effects of the initial post-processing steps. On the other hand, the last- and ext-SQD steps show impressive convergence to the reference value across all layouts/modes. For the last-iteration we register a difference of $0.47-0.58$ kcal/mol from the HCI value. The largest discrepancy is observed for Layout 2 where the two execution modes differ by $0.02$ kcal/mol for EtOH$_{1.2}$.
At the ext-SQD stage, these differences become negligible: for EtOH$_{Eq}$ in Layouts 1 and 3, parallel and serial deviations differ by just 0.001 kcal/mol, a value well below chemical accuracy and within statistical uncertainty.
This progressive convergence confirms that the SQD workflow effectively mitigates the stochastic variability and potential cross-talk effects observed in the first iteration, ensuring an accurate estimation of the ground-state energy. Taken together, our results exhibit spread distributions in the computed ground-state energies across all three layouts (Figure \ref{fig:Figure3}), which arises from the inherent stochastic nature of the quantum measurements and the following post-processing procedure. This stochasticity does not preclude the presence of hardware cross-talk, but rather masks it. However, the proposed multi-programming framework remains robust. This is because we employ SQD and ext-SQD for configuration recovery, ensuring the protocol reliably converges to the reference values for both single and parallel executions. Consequently, the approach is reliable and directly enables a more efficient utilization of QPU resources. By combining multiple independently transpiled LUCJ circuits, these parallel jobs achieve the same wall-clock time as single-circuit run while simultaneously delivering multiple accurate energy evaluations.

\section{Limitations and Outlook}

In this work, we addressed a potential operational bottleneck tied to the increased, but still limited, access quotas of cloud-based quantum computing. We exploited the Qiskit Experiments ecosystem \cite{QiskitExperiments}, and implemented a new \verb|Experiment| class tailored for chemistry-oriented problems. This new approach moves beyond the sequential job execution and enables the simulation of multiple LUCJ circuits combined with ad-hoc post-processing. Converged ext-SQD estimates differ by as little as 0.001 kcal/mol between parallel and serial executions, providing a robust and practically relevant strategy for improving QPU throughput. These results show that, despite the presence of cross-talk and stochastic variability, our approach can reliably reproduce the ground-state energy by successfully mitigating these effects.
This is especially advantageous for users operating under limited access to quantum hardware, where execution budgets represent a dominant limitation. Here, we analyzed two simultaneous LUCJ circuits, as this represents the largest system with clear advantage on quantum hardware over uniform sampling. However, our framework supports the concurrent execution of more circuits, but this is constrained by the number and quality of available qubits, as well as by the requirement that all the circuits used the same number of measurements, set to the highest value needed by any individual circuit. Additionally, while the variability at the first-SQD iteration has been attributed here to intrinsic stochasticity of QPU and SQD sampling, underlying hardware-level contributions (e.g., readout errors, gate fidelities, and coherence times) have not been directly addressed. As larger and more homogeneous QPUs become available, the framework is expected to extend naturally to broader quantum chemistry problems. In particular, it can be applied to alchemical free energy protocols \cite{BazayevaQuantumAFE} and fragmentation methods for which a custom compiler can be implemented. \cite{QuantumProtein,wang2025samplebasedquantumdiagonalizationparallel}

\begin{acknowledgement}

The authors gratefully acknowledge financial support from the National Science Foundation (NSF) through CSSI Frameworks Grant OAC-2209717 and from the National Institutes of Health (Grant Numbers GM130641). The authors also thank Danil Kaliakin for his feedback during the development and evaluation of this work.

\end{acknowledgement}

\subsection{LLM Statement}

AI writing assistance (Claude Sonnet 4.6) was used for linguistic refinement and rephrasing of the manuscript. The code developed in this study is solely the authors' work.

\section*{Code Availability}
List of the GitHuB repositories of the codes used in the present work:\\
Qiskit Experiment:\\
\url{https://github.com/Qiskit-Community/qiskit-experiments}
\url{https://qiskit-community.github.io/qiskit-experiments/index.html}

Qiskit and Qiskit IBM Runtime:\\
\url{https://github.com/Qiskit/qiskit}
\url{https://github.com/Qiskit/qiskit-ibm-runtime}

ffsim:\\
\url{https://github.com/qiskit-community/ffsim}

Qiskit SQD addon:\\
\url{https://github.com/Qiskit/qiskit-addon-sqd}

Dice solver:\\
\url{https://github.com/sanshar/Dice}

SBS solver:\\
\url{https://github.com/r-ccs-cms/sbd}

PySCF:\\
\url{https://github.com/pyscf/pyscf}

PyCI:\\
\url{https://github.com/theochem/PyCI}

The tutorial demonstrating full SQD workflow is available below:\\
\url{https://qiskit.github.io/qiskit-addon-sqd/tutorials/01_chemistry_hamiltonian.html}

\section*{Competing Interest}
The authors declare no competing interest.


\bibliography{AFE_QPU}

\newpage
\begin{center}
\section*{Supplementary Information}
\end{center}

\subsection{Ancilla Count Benchmarking}

\begin{figure}
    \centering
    \includegraphics[width=1\linewidth]{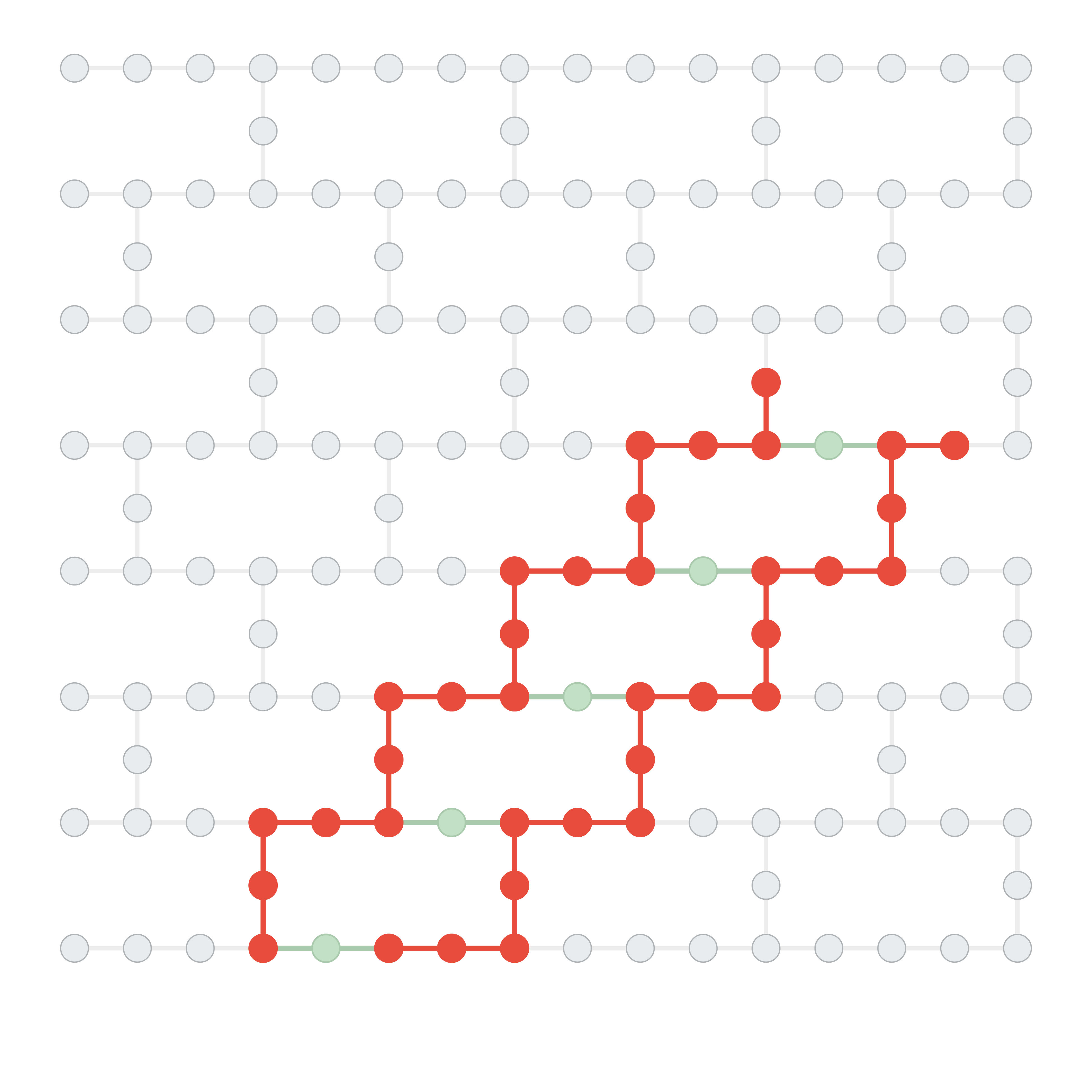}
    \caption*{Figure S1:
Quantum circuit used to benchmark the optimal number of ancillary qubits using for the  EtOH$_{Eq}$ structure. The circuit was executed on the QPU using 200,000 quantum shots. Active qubits are shown in red, while green nodes represent the ancillas evaluated across the benchmark.}
\renewcommand{\figure}{S1}
\end{figure}

\begin{table}[H] 
\centering
\begin{tabular}{cccc} 
\toprule
Ancillas Number& $|$HCl – First-SQD$|$ & $|$HCl – Last-SQD$|$ & $|$HCl – ext-SQD$|$ \\
\midrule
\multicolumn{4}{c}{1K SQD Samples} \\
\midrule
2& 12.04& 3.28& 0.24\\
3& 10.90& 3.34& 0.21\\
4& 11.02& 3.19& 0.20\\
5& 10.26& 3.10& 0.19\\
\midrule
\multicolumn{4}{c}{3K SQD Samples} \\
\midrule
2& 6.60& 0.60& 0.16\\
3& 6.40& 0.56& 0.17\\
4& 6.04& 0.62& 0.17\\
5& 6.20& 0.54& 0.16\\
\end{tabular}
\caption{Absolute energy deviation from the HCI reference at the first-, last-, and ext-SQD steps for EtOH$_{Eq}$ as a function of the number of ancilla qubits, evaluated at 1K and 3K SQD samples per batch. Results were obtained from a single execution of 200,000 quantum shots. Based on these results, 3 ancilla qubits and 3K samples were selected as the optimal configuration, balancing accuracy and computational cost, and used throughout the main text benchmarks. All values in kcal/mol.}
\label{tab:S2}
\end{table}
\newpage
\subsection{EtOH$_{1.2}$ SQD  sampling convergence}
\begin{figure}[H]
    \centering
     \includegraphics[width=0.9\textwidth, height=0.8\textheight, keepaspectratio]{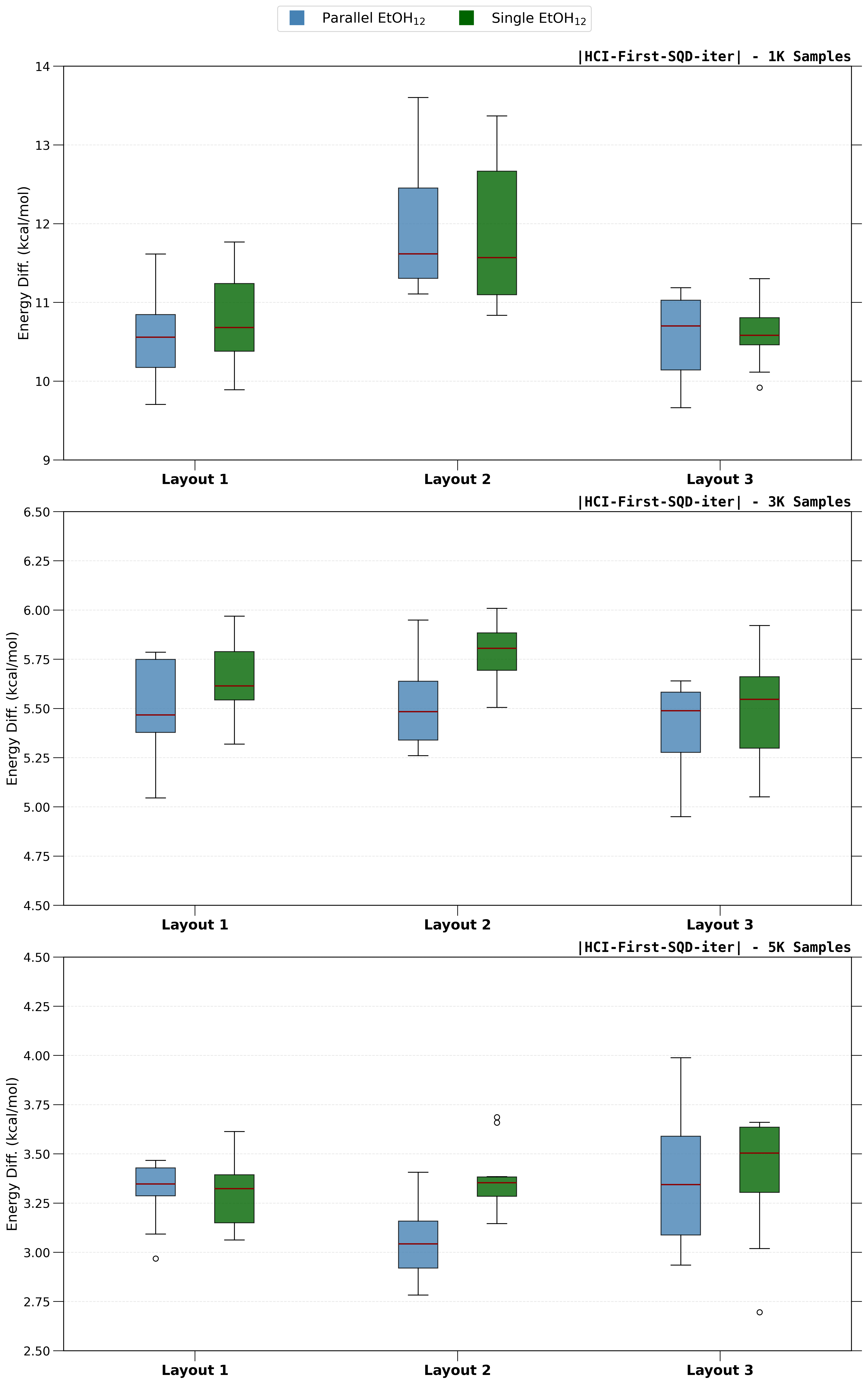}
    \caption*{Figure S2:
Distributions of the absolute energy deviation from the HCI reference at the First-SQD-iter stage for EtOH$_{1.2}$ at 1K, 3K, and 5K SQD samples per batch, across the three layouts and both execution modes over 10 independent replicates. The progressive narrowing of the distributions with increasing sample size illustrates the sampling convergence behavior of this far-from-equilibrium geometry. The 3K regime used in the main text represents a deliberately challenging but realistic sampling condition.}
\renewcommand{\figure}{S2}
\end{figure}

\end{document}